\documentclass[
 aip,
 amsmath,amssymb,
 reprint
]{revtex4-2}

\usepackage[T1]{fontenc}
\usepackage{graphicx}
\usepackage{mathrsfs}
\usepackage{xcolor}
\usepackage{textgreek}
\usepackage[hidelinks]{hyperref}
\usepackage{subcaption}
\usepackage[utf8]{inputenc}
\usepackage[english]{babel}
\newcommand{\anisoap}{\texttt{AniSOAP}}
\newcommand{\aaA}{\textalpha,\textalpha-A}

\begin{document}

\begin{abstract}

Physical vapor deposition can produce organic glasses with enhanced stability and pronounced structural anisotropy (layering) even when the constituent molecules are nearly spherical. However, the microscopic origin of the layering in spherical molecular glass formers remains unclear. We investigated vapor-deposited films of 9-(3,5-di(naphthalen-1-yl)phenyl)anthracene (\aaA) by grazing-incidence wide-angle X-ray scattering, atomistic molecular dynamics, and coarse-grained simulations. Experimentally, the layering signal varies nonmonotonically with deposition temperature and is strongest near 0.8$T_g$. Simulated scattering calculations reproduce the principal experimental features and show that the layering peak does not originate from shifts in single molecule degrees of freedom. In fact, the simulated deposited and bulk glasses show largely the same single-molecule conformations. Interestingly, a one-site coarse-grained liquid with isotropic interactions develops a comparable interfacial layering signature. These results indicate that layering in \aaA\ arises primarily from collective density correlations generated at the free surface during deposition rather than from conformational bias or orientational ordering. The findings connect structural anisotropy in vapor-deposited molecular glasses to the generic interfacial organization of simple liquids.

\end{abstract}

\title{Intermolecular and Interfacial Origins of Layering in a Near-Isotropic Organic Glass Former}

\author{Charles H. Carroll}
\affiliation{Department of Chemical and Biological Engineering, University of Wisconsin, Madison, WI, USA}
\author{Juliana I. Bonilla-Lugo}
\affiliation{Department of Chemistry, University of Pennsylvania, Philadelphia, PA, USA}
\author{Arthur Y. Lin}
\affiliation{Department of Chemical and Biological Engineering, University of Wisconsin, Madison, WI, USA}
\author{Richard Stephens}
\affiliation{Department of Chemistry, University of Pennsylvania, Philadelphia, PA, USA}
\author{Peng Luo}
\altaffiliation{Institute of Physics, Chinese Academy of Sciences, Beijing, CHN}
\affiliation{Department of Chemistry, University of Pennsylvania, Philadelphia, PA, USA}
\author{Sarah E. Wolf}
\affiliation{Department of Chemistry, State University of New York Cortland, Cortland,
NY, USA}
\author{Cindy Yueli Chen}
\altaffiliation{Evonik, North Brunswick, New Jersey, USA}
\affiliation{Department of Chemistry, University of Pennsylvania, Philadelphia, PA, USA}
\author{Dong Hyup Kim}
\altaffiliation{Evonik, North Brunswick, New Jersey, USA}
\affiliation{Department of Chemistry, University of Pennsylvania, Philadelphia, PA, USA} 
\author{Yu Wu}
\affiliation{Department of Chemistry, University of Pennsylvania, Philadelphia, PA, USA}
\author{Patrick J. Walsh}
\affiliation{Department of Chemistry, University of Pennsylvania, Philadelphia, PA, USA} 
\author{Patryk W\k{a}sik}
\affiliation{National Synchrotron Light Source II, Brookhaven National Laboratory, Upton, NY, USA}
\author{Whitney S. Loo}
\affiliation{Department of Chemical and Biological Engineering, University of Wisconsin, Madison, WI, USA}
\author{Zahra Fahkraai}
\affiliation{Department of Chemistry, University of Pennsylvania, Philadelphia, PA, USA}
\author{Rose K. Cersonsky}
\altaffiliation{rose.cersonsky@wisc.edu}
\affiliation{Department of Chemical and Biological Engineering, University of Wisconsin, Madison, WI, USA}
\affiliation{Department of Materials Science and Engineering, University of Wisconsin, Madison, WI, USA}
\affiliation{Data Science Institute, University of Wisconsin, Madison, WI, USA}

\flushbottom
\maketitle
\thispagestyle{empty}

\section{Introduction}

Organic light-emitting diodes (OLEDs) are critical components in modern phones and consumer electronics. Small organic molecules, which transfer negative and positive charges to the emissive layer, are used to create electron and hole transport layers in OLEDs.\cite{jikang_applicatioin_2025} This charge transport is governed not only by the single molecule electronic structure but also by the alignment of molecules relative to each other, where strong alignment can facilitate orders of magnitude increases in charge transport fluxes.\cite{yokoyama_molecular_2011,gujral_anisotropic_2018} Strong molecular alignment often manifests as anisotropy, or the dependence of the structure and performance on the direction of the material. Thus, understanding the fundamental physics driving this anisotropy could be used to more effectively engineer OLEDs.

Physical vapor deposition (PVD) is a popular method to produce anisotropic glass films with enhanced electronic properties.\cite{luo_surface_2023,anisotropicSG_review} 
In PVD, evaporated molecules are deposited onto a heated substrate (T$_\text{dep}$).
The presence of this substrate has also been connected to orientational order in films made of nonspherical molecules\cite{gujral_structural_2015,gujral_anisotropic_2018,lyubimov_orientational_2015} and even phase separation.\cite{alahe_kinetically_2026} 
The process leaves a heated free surface exposed to vacuum, where the molecules can re-organize differently than in the bulk.\cite{sg_physics} 
Simulation work has shown that layers can form at the free surface\cite{hoang_free_2011} and that the orientation of nonspherical molecules is a feature of equilibrium supercooled liquids,\cite{dalal_tunable_2015,bishop_vapor_2019} suggesting that anisotropic order could arise from equilibrium effects that are buried during deposition.
Hence understanding the general trends in how molecular shape and deposition conditions affect the resulting structure of the film has been a central goal of the field.\cite{anisotropicSG_review,bishop_sg_ratetempsuperposition,aaA_Zhang_THEpaper}

\begin{figure}[ht]
\includegraphics[width=0.4\linewidth]{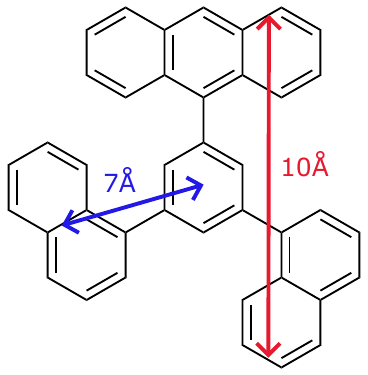}
\caption{The chemical structure of \aaA\ with characteristic dimensions shown.\vspace{-12pt}}
\label{fig:aaA_struc}
\end{figure}


\begin{figure*}[ht]
\includegraphics[width=0.8\linewidth]{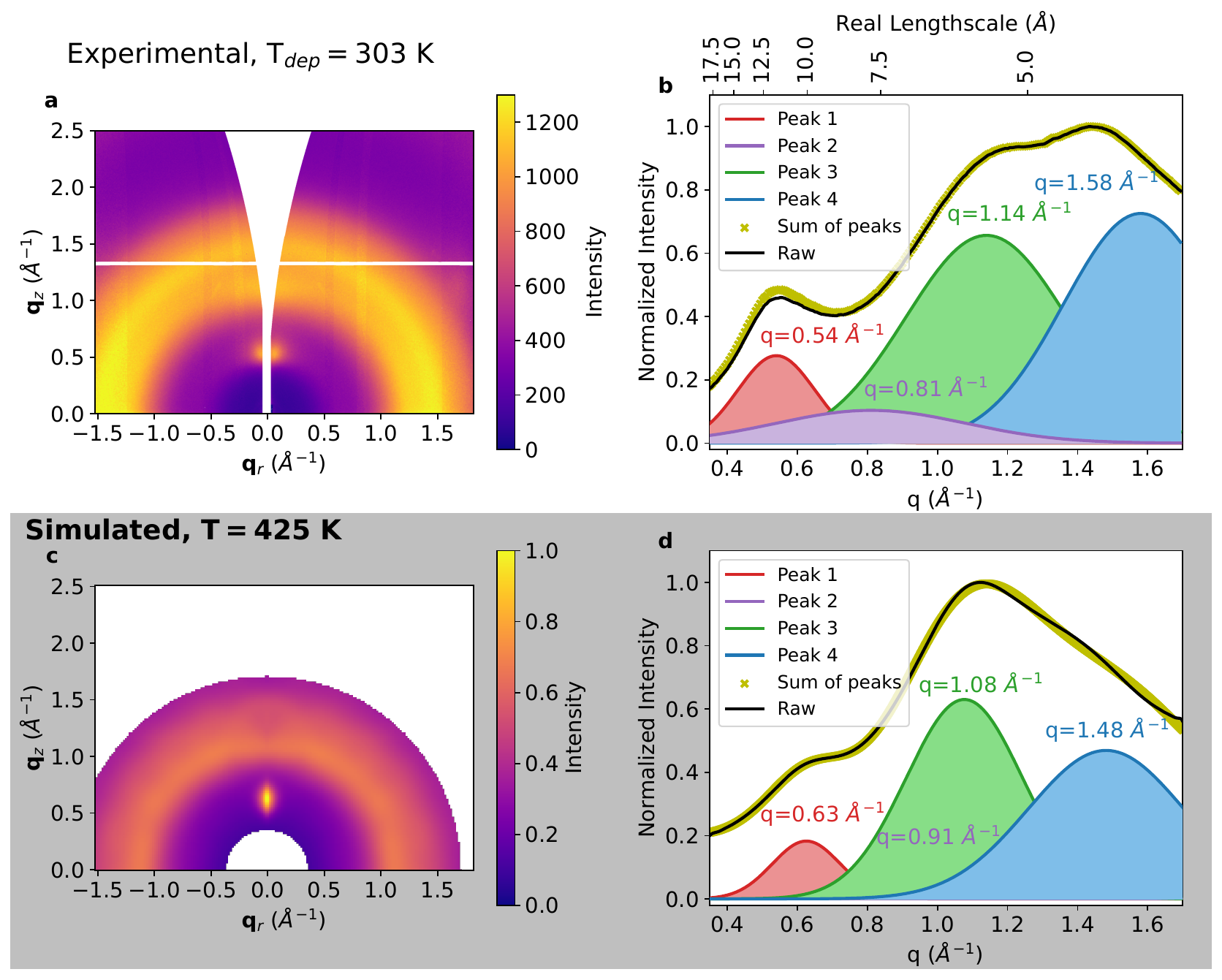}
\caption{X-ray scattering results of \aaA. (a) Experimental 2D GIWAXS image showing the $q_z$ vs $q_{r}$ of a vapor-deposited 310 nm film of \aaA\ at $T_{\text{dep}}=303$ K. (b) Normalized 1D intensity averaged over $\Psi$ as transformed by Equation \ref{eq:giwaxs_total_model} for a 310 nm film of \aaA\ at $T_{\text{dep}}=303$ K. (c) Simulated 2D GIWAXS image calculated following Equation \ref{eq:Irz}. (d) Average of the simulated GIWAXS image over $\Psi$.}
\label{fig:int_grid}
\end{figure*}

Still, much of this work still does not explain the formation of layered structures in near-spherical molecules, which do not exhibit the orientational preferences seen in highly-anisotropic glass formers. Here, we combine experiment and simulation to investigate the deposited structure of a model isotropic glass former, 9-(3,5-di(naphthalen-1-yl)phenyl)anthracene~(\aaA), $T_\text{g}=364$ K, shown in Figure \ref{fig:aaA_struc}. This molecule possesses equal length and width, but the steric repulsion of the large aromatic groups produce frustrated intermolecular interactions. Photoluminescence experiments have shown that this molecule behaves isotropically.\cite{liu_birefringent_2017} However, in PVD films of \aaA\, spherical ellipsometry\cite{liu_birefringent_2017} and grazing incident X-ray scattering experiments reveal optical anisotropy\cite{aaA_Zhang_THEpaper}, showing that birefringence can be due to orientation effects or due to packing in the process of PVD.

We seek to determine how the intramolecular degrees of freedom and center of mass ordering interact to create molecularly layered films. To do so, we analyze the structure of glasses deposited at different temperatures using grazing incidence wide angle X-ray scattering (GIWAXS), which provides nanometer level resolution of the experimental structure which we compare with molecular dynamics (MD) simulations at the all-atom and coarse-grained level. From these simulations, we compute how the interactions at different length scales contribute to the structure.

Our results indicate that the layering of spherical molecules does not impose detectable shifts in the single molecule conformations, and that the total structure is well described by intermolecular correlations of the functional groups and liquid-like centers of mass ordering. We test this finding by comparing results against a highly-coarse-grained equilibrium model of the surface, and find surprisingly good agreement. This result extends previous understanding of the enhanced surface mobility of deposited glasses by identifying an underlying thermodynamic minimum based on a highly general liquid model.

\section{Results}

\begin{figure*}[ht]
\includegraphics[width=0.8\linewidth]{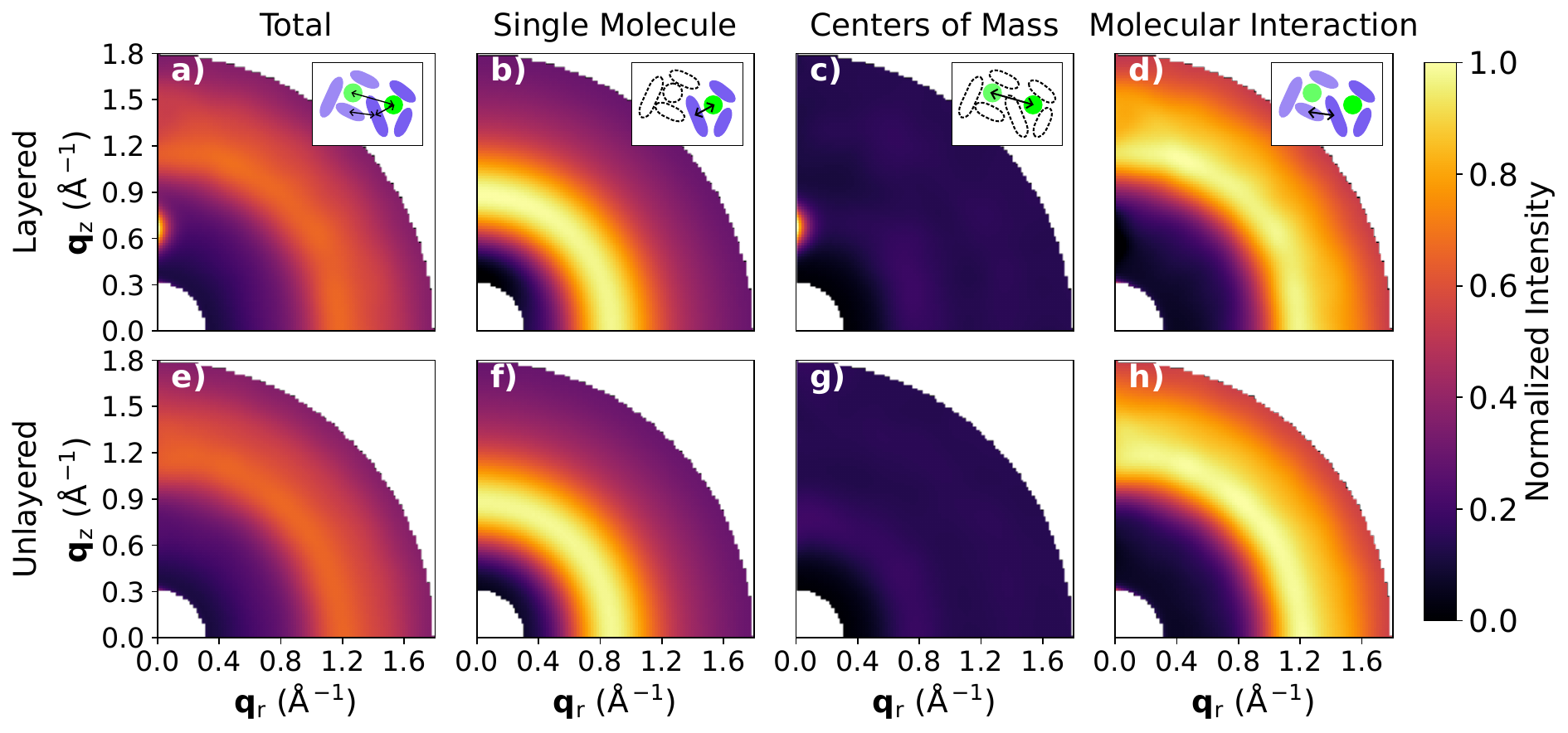}
\caption{Decomposition of the simulated two-dimensional structure factor calculated following Equation \eqref{eq:Srz} for layered and unlayered structures. The upper and lower rows show averages over simulations classified as layered and unlayered, respectively. The columns show (a,e) the total structure factor, (b,f) the ensemble-averaged single-molecule contribution, (c,g) the center-of-mass structure factor, and (d,h) the structure factor representing molecular interactions. Inset schematics visualize the physical meaning of each component. The z-axis is defined normal to the layers. Plots b and f are adjusted following Equation \eqref{eq:finite_size_correction} to remove the exponential low $q$ behavior for finite sized systems. Each column is normalized independently to emphasize the position and angular distribution of the structure factor features; intensities must therefore not be compared quantitatively between columns.
}
\label{fig:structure_factor}
\end{figure*}

\subsection{Structural features of the experimental scattering signal}

The experimental GIWAXS intensity plot shown in Figure \ref{fig:int_grid}a, corresponding to $T_\text{dep}$ = 303~K = 0.83$T_\text{g}$, reveals two prominent signals. The first is the halo at $q=$1.1-1.4 \AA$^{-1}$. The lack of direction dependence indicates that the structural features are isotropic. The real length scale of this signal is about 4.5-5.7 \AA, which is the characteristic length scale of $\pi-\pi$ interactions. The second is the strong signal at $q_z=$0.5-0.6 \AA$^{-1}$ concentrated on the $q_z$ axis. This anisotropic signal represents strong ordering orthogonal to the substrate at a length scale of about 10-12 \AA. The similarity of this distance with the molecular size shown in Figure \ref{fig:aaA_struc} indicates the molecules of \aaA\ form sheets across the substrate with the sheets then stacked on top of each other. This phenomenon is also known as layering, with the anisotropic peak in the GIWAXS signal termed the ``layering peak.'' Similar layering peaks have been observed in vapor deposited films of spherical and nonspherical molecules. \cite{aaA_Zhang_THEpaper,bagchi_over_2020,bagchi_origin_2019,bishop_sg_ratetempsuperposition}

The two dimensional (q$_r$,q$_z$) intensity plot is projected across the scattering vector magnitude and the angle in between the scattering ($\Psi$) to extract the intensity averaged over $\Psi$ plot against $q_\rho$ as shown in Figure \ref{fig:int_grid}(b). We fit the projected signal over the range $q_{\rho}$ 0.3-1.8 \AA$^{-1}$ using four Gaussian distributions, following Equation \eqref{eq:giwaxs_total_model}. We identify four peaks: Peak 1, 2, 3, and 4 - where Peak 1 is at the smallest scattering vector magnitude corresponding to the layering process up to T$_\text{dep}\approx$ 0.9T$_g$. Peak 2 has a weak and broad intensity with regards to the magnitude of the wavevector, but has a high angular character (see Section I B in the supplementary material). Finally, Peaks 3 and 4 comprise the isotropic halo.

Then, we calculated the X-ray scattering intensity plots from the all-atom simulations following Equation \eqref{eq:Irz}. The all-atom configurations are generated using umbrella sampling where a variable strength force drives the simulation towards layered structures (see Section \ref{sec:molecular_dynamics} in \emph{Methods}). The simulations were run at 425 K to compensate for the higher glass transition temperature (about 500 K, see Figure S3) associated with the shorter time scale of simulation.\cite{madanchi_is_2025} The result shown in Figure \ref{fig:int_grid}(c) demonstrates a similar layering peak and isotropic halo. The one dimensional plot in Figure \ref{fig:int_grid}(d) allows a quantitative comparison. We again fit four Gaussian distributions restricting the position of each peak to be within 0.1 \AA$^{-1}$ of the experimental peaks. There are three discrepancies visible:  Peak 1 shifted to a higher wavevector magnitude in simulation, indicating the simulation layers are about 14\% closer than in experiment; Peak 2 is not visible in simulation; Peak 4 is significantly broader in simulation, suggesting the short range ordering of the aromatic groups is somewhat less regular in simulation than in experiment. This is likely due to the higher temperature and limited time for relaxation in simulation, although there can be some uncertainties attributed to the force-field parameterization. Nonetheless, the layering peak is visible and still very close to the characteristic size of \aaA\ and Peak 3 agrees quite well between experiment and simulation. We therefore consider the simulation to be an adequate representation of the atomic structure of the film, and accurate enough to capture the general physics of the situation.

\subsection{Separation of degrees of freedom}

\begin{figure}[ht]
\centering
\includegraphics[width= \linewidth]{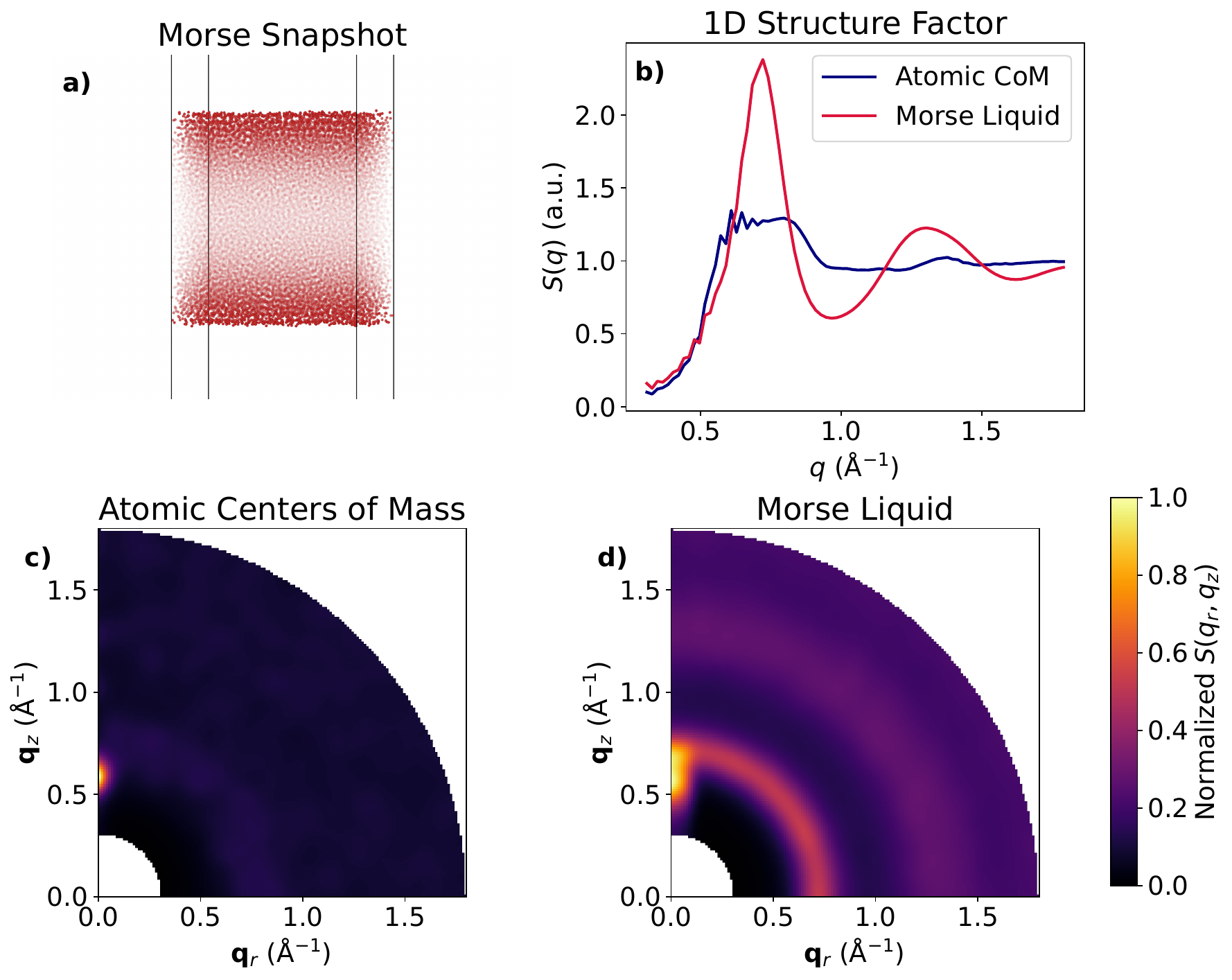}
\caption{Summary of coarse-grained results. (a) Snapshot of Morse simulation, with a transparency gradient to emphasize the surfaces in the visualization. (b) Structure factor intensity as a function of total wavevector magnitude. (c) Structure factor intensity plot of the centers of masses calculated from layered all-atom simulation. (d) Structure factor intensity plot of the Morse liquid.}
\label{fig:com_morse_comparison}
\end{figure}
Next, we consider the contributions of the various molecular degrees of freedom to the total structure by using the structure factor of Equation \eqref{eq:Srz}. This equation is closely related to the X-ray scattering signal of the previous section but without the added complication of $q$-dependent atomic form factors. Using simulation data, we calculated the structure factor of various subsets of the atoms in the simulation. The resulting signal is shown in Figure \ref{fig:structure_factor}, where we separate the simulation data into layered (a,b,c,d) and unlayered (e,f,g,h) sets. Simulations with a layering peak of at least 60\% of the isotropic halo intensity are classified as layered. We calculated the total structure factor (a,e), the average single molecule structure factor (b,f), and the structure factor of the molecular centers of mass (c,g). The convolution of the single molecule and center of mass structure factors is an approximation of the total structure factor that neglects the correlations between the conformations of neighboring molecules. In other words, it represents a structure with the same center of mass positions but with noninteracting molecules. The difference between the total structure factor and the noninteracting convolution we term the ``molecular interaction'' structure factor,  shown in Figure \ref{fig:structure_factor}(d,h). This is the excess correlation arising from the interactions between nearby molecules.

The isotropic halo of the total structure factor comes from the correlations between molecules. The observation that the halo in the single molecule structure factor in Figure \ref{fig:structure_factor}(b,f) is at a larger length scale (0.9 \AA$^{-1}$) than the molecular interaction halo in Figure \ref{fig:structure_factor}(d,h) (1.2 \AA$^{-1}$) indicates that the functional groups of neighboring molecules approach closer to each other than the groups on the same molecule (Figure \ref{fig:structure_factor}(b,f)). We also observe that the halo in the total structure factor (1.2 \AA$^{-1}$, Figure \ref{fig:structure_factor}(a,e)) matches that of the molecular interactions. This suggests that the molecular level structure is based on the interlocking of the molecular functional groups, rather than each \aaA\ molecule taking up their own roughly spherical volume and interacting at the edges of their spheres. This implies that a key barrier for relaxation is the steric frustration of the enmeshed structure. In addition, the isotropic halo in the experimental GIWAXS signal corresponds to this length scale, especially Peak 3 (1.14 \AA$^{-1}$, Figure \ref{fig:int_grid}(b)), and would therefore appear to be a direct experimental probe into the tangled structure of the film.

Comparing the layered and unlayered structure factors, the single molecule (Figure \ref{fig:structure_factor}(b,e)) and molecular interaction structure factors (Figure \ref{fig:structure_factor}(d,h)) are similar between them. The layering signal arises entirely from the ordering of the centers of mass (Figure \ref{fig:structure_factor}(c,g)). It may be intuitive to suppose that layering would be accompanied by a change in single molecule conformation or orientation. Perhaps layering accesses an entirely different, more stable molecular level organization. Indeed, for nonspherical molecules, layering is strongly associated with the onset of ensemble level orientational order. \cite{lyubimov_orientational_2015,walters_influence_2017,ferron_characterization_2022,dalal_tunable_2015} For \aaA\ however, we observe no such trend, which would appear as an anisotropic signal in the single molecule structure factor. The lack of orientational order also agrees with evidence from spherical ellipsometry measurements of \aaA,\cite{liu_birefringent_2017,aaA_Zhang_THEpaper} which implies the layering mechanism is at a larger length scale. It also suggests that the molecular degrees of freedom could be accounted for in a simplified isotropic effective interaction. This interpretation is also supported by the use of machine learning molecular descriptors, see Section IV in the supplementary material.

\subsection{Mapping to equilibrium simple liquids}

Given the significance of the equilibrium interfacial liquid identified in the literature\cite{singh_molecular_2011,bagchi_over_2020,bishop_vapor_2019} and the results of the all-atom structure factor analysis, we consider a simplified coarse-grained model where each molecule is represented by a single isotropic bead. This approach differs from the coarse-graining approach most commonly used in the field, where a small number of beads are used to roughly approximate a molecule's shape.\cite{bagchi_over_2020,dalal_tunable_2015,walters_influence_2017,lyubimov_orientational_2015, aaA_Zhang_THEpaper} We justify the complete neglect of molecular shape by appealing to the isotropy of the single molecule and molecular interaction structure factors (Figure \ref{fig:structure_factor}(b,d,f,h)). 

Determining the optimal coarse-grain interactions for a given system is a complex and open problem. However, as we are interested in more general physical trends, we avoid this problem by simply choosing a pairwise Morse potential for our particles due to its softness and use in highly coarse-grained systems. \cite{zhang_parameterizing_2014,liew_coarse_2003,chiu_coarse-grained_2010}
This simplified model potential is meant to probe the applicability of general liquid physics to the deposited \aaA\ glass, rather than the role of specific features of the molecule.

For our simplified equilibrium model, we therefore simulate a Morse liquid of 25,000 particles exposed to vacuum at the each interface (see Section \ref{sec:molecular_dynamics} in \emph{Methods}).
The comparison between the one dimensional structure factors for the centers of mass from the all-atom simulation and the particles in the Morse simulation, shown in Figure \ref{fig:com_morse_comparison}(b), shows that the Morse liquid structure factor is more strongly peaked than that of the atomistic centers of mass. The Morse potential in this case is more repulsive at short distances than the actual effective interaction. The intensity plots of Figures \ref{fig:com_morse_comparison}(c,d) reveal the presence of the layering peak in the all-atom and Morse liquid simulations. A discrepancy between the two simulations is the prominence of the isotropic halo in the Morse liquid. This halo represents the underlying bulk liquid ordering, which corresponds to the oscillations of the radial distribution function seen in liquids. A similar feature can be observed in the all-atom structure factor although much fainter and broader, indicating liquid-like ordering is weaker in the all-atom system. The short range repulsiveness of the Morse potential leads to the narrowness of the halo in the liquid and the presence of many particles in the bulk below the surface increases the relative signal of the bulk structure. Importantly, that we observe a strong layering peak even with a simple untuned potential is strong evidence that the center of mass ordering of a deposited glass film comes from a more general phenomenon of simple liquids.

\begin{figure}[ht]
\centering
\includegraphics[width=0.9\linewidth]{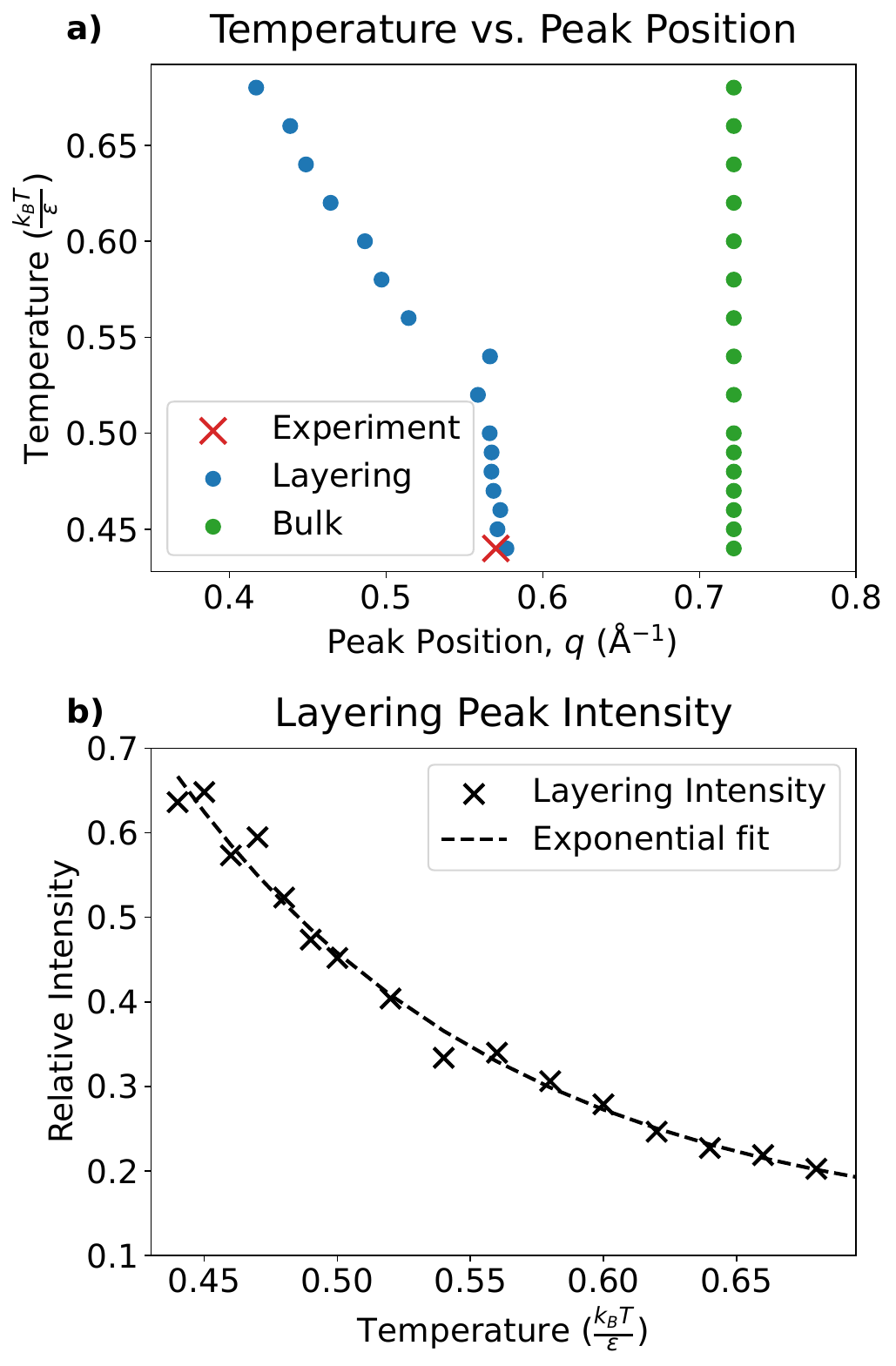}
\caption{Temperature dependence of peaks for the Morse liquid. (a) The position of the simulation surface and bulk peaks with respect to temperature. Peaks 1 and 2 are marked with filled circles, with colors matching those of Figure \ref{fig:int_grid}(b). (b) The ratio of the layering peak intensity over the bulk peak intensity as a function of deposition temperature. The black crosses are results from simulation of the Morse liquid with a free surface. The simulation temperatures are scaled so that the minimum non-crystallizing temperature maps to the lowest experimental temperature with a layering peak, marked with a red cross in (a).}
\label{fig:temp_dependence}
\end{figure}

To understand this phenomenom in the simplified model, we consider the temperature dependence of the structure factor peaks from the coarse-grained equilibrium model in Figure \ref{fig:temp_dependence}. The peak corresponding to ordering in the bulk (green), which arises from the isotropic ordering of the Morse liquid far from the surface, shows no temperature dependence over the examined range. In contrast, the layering peak (blue) is strongly dependent on the temperature. This peak corresponds to the liquid structure at the surface, and moves to lower wavevector magnitudes as the temperature increases. Figure \ref{fig:temp_dependence}b shows the relative intensity of the layering signal with respect to the bulk signal decays exponentially with temperature. The shift of the layering peak to longer length scales and the decrease in intensity indicates the breakdown of the layered structure at high temperatures, with only the bulk liquid structure commonly associated with liquids remaining. This temperature dependence in the Morse model suggests that the high temperature sensitivity of vapor deposited glasses may similarly correspond to the sensitivity of the underlying equilibrium simple liquid surface.

\section{Discussion}


Our results suggest that layering in PVD glass films is the result of the molecular center of masses interacting as an equilibrium liquid. We demonstrate that a simple equilibrium coarse-grained model is able to replicate key features of the all-atom simulation (Figure \ref{fig:com_morse_comparison}) and demonstrate a surface structure highly sensitive to temperature (Figure \ref{fig:temp_dependence}). We propose, based on structure factor analysis of the all-atom simulation data (Figure \ref{fig:structure_factor}) and experimental GIWAXS data (Figure \ref{fig:int_grid}) that the molecular interactions are mediated by the flexible aromatic groups of \aaA\ resulting in a simple, isotropic effective potential. Therefore we conclude that the layering in \aaA\ deposited films is closely related to a general equilibrium phenomenon of simple liquids and this effect is likely general to other molecular PVD films.


Our results build on previous work concerning the origin of the anisotropic structure and increased stability of PVD glasses. 
Enhanced mobility has long been used to explain PVD glass's enhanced stability,\cite{swallen_organic_2007} although it has been observed that greater surface diffusion does not necessarily lead to more stable glasses, such as in \aaA\ \cite{samanta_exploring_2019} or in polymeric glasses.\cite{chai_direct_2014} The equilibrium surface has been empirically connected to molecular layering,\cite{singh_molecular_2011} although the reason why this kind of center of mass ordering is the relevant thermodynamic minimum across a wide variety of molecular glasses has not been elaborated upon. Our results suggest that the liquid-like ordering at the surface is a prominent thermodynamic driving force for many small organic glass formers.


We expect these results to be applicable to PVD glasses with mostly spherically shaped molecules, with deviations expected when effective interactions become directional, such as with molecular anisotropy or strong intermolecular interactions \cite{tylinski_vapor_2016,moore_effects_2019}. Thus, while the layering effects in isotropic molecules parallel those of non-spherical molecules, the causal mechanism differs.

\section{Conclusion}

In this study, we combined grazing-incidence X-ray scattering with atomistic and coarse-grained simulations to identify the microscopic origin of the molecular-scale layering in vapor-deposited \aaA. Our analysis separates translational layering from molecular orientation and intramolecular conformation. We find that the experimentally observed anisotropic structure factor feature is primarily associated with intermolecular center-of-mass correlations and does not require orientational or conformational ordering. Our results are consistent with previous work showing that the unique properties of the deposited glass arise from surface equilibration and subsequent burying, with the temperature sensitivity coming from the careful balance of surface mobility and thermodynamics.\cite{thoms_in_2020,berthier_origins_2017,rodriguez_surface_2019} 


The various features of molecular layering have become well understood in recent years, but the molecular organization within the film and its origin has remained unclear. Our work provides evidence that molecular conformations and interactions need not significantly change from bulk behavior to allow for layering. This work suggests that the intramolecular degrees of freedom mediate simple interactions between the molecules and that the layering of \aaA\ is a result of the molecular centers of mass behaving as an equilibrium simple liquid at the surface. This finding suggests that the physical driving force of glass morphology entails a competition between surface effects (which would favor layering) and the frustration of local molecular interactions.

\section*{Supplementary Material}
\noindent The supplementary material contains further details on GIWAXS measurements, umbrella sampling protocols, and additional characterization of the simulated all-atom results using the \anisoap\cite{lin_expanding_2024, jorgensen_interpretable_2026} methodology and density calculations.

\section*{Acknowledgments}
\noindent This research was primarily supported by NSF through the University of Wisconsin Materials Research Science and Engineering Center (DMR-2309000). This investigation used the 12-ID beamline of the National Synchrotron Light Source II, a US Department of Energy (DOE) Office of Science User Facility, operated for the DOE Office of Science by Brookhaven National Laboratory under contract no. DE-SC0012704. This work used the Anvil computing center at Purdue University through allocation CHM250054 from the Advanced Cyberinfrastructure Coordination Ecosystem: Services \& Support (ACCESS) program, which is supported by U.S. National Science Foundation grants no. 2138259, 2138286, 2138307, 2137603, and 2138296. Thanks to M. D. Ediger for insightful discussions and Dr. Po-Cheng Kung for insights in interpreting simulation structure factor results.

The funders had no role in study design, data collection and
analysis, decision to publish, or preparation of the manuscript.

\section*{Author Declarations}

\subsection*{Conflict of Interest}
\noindent The authors have no conflicts to disclose.

\subsection*{Author Contributions}
\noindent CC: Conceptualization, Data curation, Formal analysis, Investigation, Methodology, Software, Validation, Visualization, Writing - Original Draft, Writing - Review \& Editing; JB: Conceptualization, Data curation, Formal analysis, Investigation, Methodology, Validation, Visualization, Writing - Original Draft, Writing - Review \& Editing; AL: Formal analysis, Software, Visualization; RS: Resources; PL: Resources; SW: Resources; CC: Resources; DK: Resources; YW: Resources; PW: Resources; PW: Resources; WL: Formal analysis, Supervision, Writing - Review \& Editing; ZF: Conceptualization, Data curation, Formal analysis, Funding acquisition, Investigation, Methodology, Validation, Visualization, Writing - Original Draft, Writing - Review \& Editing; RC: Conceptualization, Data curation, Formal analysis, Funding acquisition, Investigation, Methodology, Software, Validation, Visualization, Writing - Original Draft, Writing - Review \& Editing.

\section*{Data Availability}
\noindent Data will be made available in a Zenodo repository upon publication.

\section{Methods}

\subsection{Synthesis and sample preparation}

\aaA\ (structure shown in Figure \ref{fig:aaA_struc}, $T_g=364\pm2$~K\cite{aaA_Zhang_THEpaper}) was synthesized via the Suzuki cross-coupling reactions and purified as detailed in previous publications.\cite{aaA_syn,aaA_lib} \aaA\ in powder form was placed in alumina crucibles and pre-melted in a vacuum oven (Fisherbrand Isotemp 281A, P $\approx 1$~kPa) at roughly $543$~K before being placed into a custom-made deposition chamber with a base pressure of $10^{-7}$~torr for thermal deposition (as detailed in earlier publications. \cite{aaA_Zhang_THEpaper,aaA_lib,aaA_syn})

Films of \aaA\ with a thickness of 310~nm were vapor-deposited onto silicon substrates (Si(100) with native oxide, Virginia Semiconductor, Inc.), which were mounted on a bridge between two temperature-controlled copper stages at two different temperatures to produce temperature-gradient (T-grad) samples. Details of the calibration and $T_\text{dep}$ determination can be found in the supplementary material and in our previous publications.\cite{aaA_Zhang_THEpaper,aaA_lib,aaA_syn} GIWAXS experiments were performed on T-grad samples held in the $T_\text{dep}$ range of 240~K$\geq T_\mathrm{dep}\geq340$~K, at a deposition rate of $r_\mathrm{dep} = 0.02 \pm 0.02$~nm/s, monitored with a quartz crystal microbalance (QCM, Inficon STM-2).

After deposition, the temperature of the hot side of each T-grad sample was rapidly quenched, and the cold side was heated to room temperature. Samples were removed and stored at roughly 254~K prior to characterization.

\subsection{Grazing Incident Wide-Angle X-ray Scattering}

GIWAXS data were obtained at room temperature using the 12-ID (Soft Matter Interfaces) beamline at the National Synchrotron Light Source II (NSLS-II) of the Brookhaven National Laboratory. A PILATUS 900 kW detector was used with a sample-to-detector distance of 277~mm. The incident angle was set at 0.1$^{\circ}$ with respect to the plane of the film. The energy of the X-ray beam was set to 16.1~keV ($\lambda = 0.77009$ {\AA}).  The beam scanned 33-grid points along the T-grad sample, with a distance of 0.2~cm between each measured point. At each point, four images were taken, with the detector being moved horizontally on a fixed arc at 0$^{\circ}$, 2$^{\circ}$, 18$^{\circ}$, and 20$^{\circ}$ relative to the incident beam, to capture a comprehensive span of scattering angles. Each image was collected with an exposure time of 0.3~s. A Python-based (version 3) package developed at NSLS-II was used to perform post-acquisition data processing. 

Figure \ref{fig:int_grid}a shows an example of the 2D GIWAXS data. We note that in some images, the integration did not fully mask defects; these features (for example, the dark stripe around $q_z$ = 1.3 \AA$^{-1}$ in Figure \ref{fig:int_grid}a), along with regions with intense scattering from the X-ray source, were masked with white stripes.

As schematically shown in Figure S1, the azimuthal angle ($\Psi$) was defined as $\Psi= 0^{\circ}$ corresponding to an out-of-plane direction ($q_z$) and $\Psi= \pm 90^\circ$ to an in-plane direction ($q_{r}$). 
A transformation to cylindrical coordinates was also performed to obtain the angular dependence of the intensity and to generate averaged intensity profiles. The cylindrical coordinate vectors $q_\rho$ and angle $\Psi$ were defined as:
\begin{equation}
    \begin{aligned}
        q_{\rho} &= \sqrt{q_{z}^2+q_{r}^2},
        \\
        \Psi &= \arctan\left(\frac{q_{r}}{q_{z}}\right).
    \end{aligned}
\end{equation}

In the cylindrical geometry, the normalized intensity was fitted to the sum of four Gaussian functions:

\begin{equation}
I_{\mathrm{fit}}(q_{\rho},\Psi)
=\sum_{k=1}^{4}
G_k\left(q_\rho-q_k, \Psi\right).
\label{eq:giwaxs_total_model}
\end{equation}
Where $I_{fit}$ is the total fitted intensity of the measurement, $G_k(q_\rho,\Psi)$ represents a Gaussian function with angular-dependent intensity. It was found that the two peaks at lower $q$ values, $G_1(q_\rho, \Psi)$ and $G_2(q_\rho, \Psi)$ had angular dependencies, while  $G_3(q_\rho, \Psi)$ and $G_4(q_\rho, \Psi)$ did not show any angular dependence, even if they were considered as anisotropic functions. As such, for simplicity, they were fitted to isotropic Gaussian functions.


\subsection{Molecular Dynamics}
\label{sec:molecular_dynamics}

We used the Optimized Potential for Liquid Simulations (OPLS) all-atom (AA) force field,\cite{jorgensen_development_1996} which has been previously used to study glass formers\cite{zhang_study_2007,siachouli_effect_2024} to perform simulations of \aaA\ in LAMMPS.\cite{thompson_lammps_2022} Parameters, including many-body interactions, were generated using the online LigParGen tool using the 1.14*CM1A partial charge model.\cite{dodda_ligpargen_2017,jorgensen_potential_2005,dodda_localized_2017}

To validate this potential, we performed umbrella sampling using LAMMPS and PLUMED\cite{bonomi_promoting_2019,tribello_plumed_2014} to compare the stable layering separation to that exhibited in GIWAXS data. Systems of 320 molecules were initially equilibrated in the NPT ensemble at 1000~K using a low target pressure to relax the simulation cell prior to cooling. The molecules were restrained to their layers using a harmonic restraint (a so-called ``bias'') to the layer plane. We then ran simulations at 350~K using an anisotropic barostat until they demonstrated 50 uncorrelated measurements of the volume across time, as calculated by pyMBAR,\cite{shirts_statistically_2008} after which we collected biased sampling data for another 0.5~ns and computed the free energy profile of the separation distances, again using pyMBAR, and as shown in Figure S2 in the supplementary material. 

Following this validation, we ran simulations with 512 molecules, again using biasing to study multiple degrees of anisotropic layering, from bulk glass (showing no layering) to systems exhibiting strong layering consistent with experiment. Following simulations at 1000~K and 1~bar, we systematically quenched our systems at a rate of 30~K/ns until reaching 425~K, slightly below the experimental glass transition temperature for this system (as calculated following the method of \citet{suter_rapid_2025} and shown in Figure S3 in the supplementary material). At 425~K, we ran simulations, again, until sampling 30 uncorrelated volumes across time, after which we removed the simulation biases and collected simulated data in the NVT ensemble for 1~ns. The results described in the main text are based on these final production runs. All simulations were run with a simulation timestep of 1 fs.

Coarse grained simulations of the \aaA\ surface were performed using a single bead for each molecule interacting via the isotropic Morse potential,
\begin{equation}
    U(r)=\varepsilon\left(1-\exp[-a(r-r_c)]\right)^2,
    \label{eq:morse}
\end{equation}
with $\varepsilon=1$, $a=3$, and $r_c=1$ chosen to approximate the softness of the \aaA\ molecule interactions. Simulations of 25,000 particles were run in the NVT ensemble at dimensionless temperatures from $T=0.02-2.0$ using OpenMM.\cite{eastman_openmm_2024} We initialized the system to resemble a dense film in contact with vacuum, consistent with PVD procedures. We then ran simulations for 50,000$\tau$. We use the last 40,000$\tau$ to calculate the structure factors of the particles along the film/vacuum interface. Equilibrated density profiles are shown in Figure S5 for $T=0.45$. Systems below $T=0.4$ were observed to crystallize.

\subsection{Simulated X-ray Scattering}

The structure factor is related to the Fourier components of the density field,
$$\rho\left(\boldsymbol{q}\right)=\int\rho_i\left(\boldsymbol{r}\right)\exp{\left(i\boldsymbol{q}\cdot\boldsymbol{r}\right)}d\mathbf{r}.$$
where the number density of species $i$, $\rho_i$, comes directly from the N particles of the simulation,
$$\rho_i\left(\boldsymbol{r}\right)=\sum_{j=1}^{N_i}\delta\left(\boldsymbol{r}-\boldsymbol{r}_j\right).$$

The structure factor, $S\left(\boldsymbol{q}\right)$, is the spatial correlation of the number density in reciprocal space,
$$S\left(\boldsymbol{q}\right)=\frac1N\sum_{i,j}\rho_i\left(\boldsymbol{-q}\right)\rho_j\left(\boldsymbol{q}\right)=\sum_{i,j}S_{ij}\left(\boldsymbol{q}\right),$$
where $S_{ij}\left(\boldsymbol{q},t\right)$ are the partial structure factors. The structure factor is often spherically averaged over all wave vectors of the same magnitudes and over time, however here we transform the data from cartesian ($q_x,q_y,q_z$) coordinates to circularly averaged spherical ($q_r,q_z$) coordinate. We use the transformed structure factor,
\begin{equation}
    S\left(q_r,q_z\right)=\frac{1}{2\pi q_r}\sum_{i,j}S_{ij}\left(\boldsymbol{q}\right)\delta_{q_x^2+q_y^2-q_r^2}.
    \label{eq:Srz}
\end{equation}
We weight the partial structure factors by the atomic form factors calculated by \citet{waasmaier_new_1995} to get the estimated X-ray scattering from simulation,
\begin{equation}
    I\left(q_r,q_z\right)=\frac{1}{2\pi q_r}\sum_{i,j}f_i\left(q\right)f_j\left(q\right)S_{ij}\left(\boldsymbol{q}\right)\delta_{q_x^2+q_y^2-q_r^2}.
    \label{eq:Irz}
\end{equation}

The two dimensional scattering and structure factor surfaces were smoothed using kernel regression as implemented in the python package \texttt{KDEpy}.\cite{white_kdepy_2023} Here, we use a Gaussian kernel of the form $K(x_1,x_2)=\exp(-(x_1-x_2)^2/2\sigma^2)$ with a $\sigma$ value of 0.08 \AA$^{-1}$. We define our coordinate system such that the z axis is always orthogonal to layering.

The low $\boldsymbol{q}$ behavior of the scattering function arises from the finite size of the system. For single molecules, this behavior can dominate the scattering and obscures the more relevant correlations between the atoms behind this finite size effect. At low $\boldsymbol{q}$ magnitudes (which correspond to length scales greater than the longest interatomic separation in the molecule), the molecule can be approximately treated as a smooth density distribution. The size effect is often accounted for by approximating this contribution as a normal distribution, \cite{brunel_in_2024,kim_comprehensive_2022}
\begin{equation}
    S_{\text{excess}}(q)=S(q)-S(0)\exp\left(\frac{-q^2R_g^2}{3}\right),
    \label{eq:finite_size_correction}
\end{equation}
where $R_g^2$ is the radius of gyration, in this case the electron weighted variance of particle positions about the center of mass, and is related to the size of the molecule. This form corresponds to the scattering contribution of a Gaussian approximation of the molecule density. 

A common method to relate single molecule structure factor to the total is to take the convolution of the single molecule factor with a lower resolution representation of the system. In Fourier space, this is quite convenient because it is simply a multiplication of the two functions. In literature, the notation often used is 
\begin{equation}
    I(q)=P(q)S(q),
    \label{eq:reciprocal_convolution}
\end{equation}
where $P(q)$ is known as the `form factor', i.e. the single particle contribution, and $S(q)$ is called the `structure factor.' \cite{pedersen_analysis_1997,hansen_theory_2013,li_small_2016} Note that we calculate the scattering signal of simulation in a similar way, except the form factor represents the electronic response of each atom and the structure factor represents each atom position. This makes reciprocal space methods well suited for investigating structure over multiple length scales.

This convolution is a physical approximation. The approximation amounts to taking the centers of mass, adding the ensemble averaged density at each center, and calculating the resulting structure factor of the smooth field. This `decoupling' approximation clearly neglects the correlation of molecular shapes between neighbors, where a molecule is not free to randomly sample the conformational ensemble, but is restricted by the conformations of its neighbors. With this in mind, we use the convolution to account for those aspects of the structure that do not depend on intermolecular correlations, and the difference between it and the total structure factor we ascribe to correlations between aromatic groups of the same molecule or adjacent molecules.

\bibliography{references}

@article{aaA_Zhang_THEpaper,
  author = {Zhang, Aixi and Moore, Alex R. and Zhao, Haoqiang and Govind, Shivajee and Wolf, Sarah E. and Jin, Yi and Walsh, Patrick J. and Riggleman, Robert A. and Fakhraai, Zahra},
  title = {The role of intramolecular relaxations on the structure and stability of vapor-deposited glasses},
  journal = {The Journal of Chemical Physics},
  volume = {156},
  number = {24},
  pages = {244703},
  year = {2022},
  month = {06},
}

@article{anisotropicSG_review,
  author = {Ediger, M. D. and de Pablo, Juan and Yu, Lian},
  title = {Anisotropic Vapor-Deposited Glasses: Hybrid Organic Solids},
  journal = {Accounts of Chemical Research},
  volume = {52},
  pages = {407-414},
  year = {2019},
}

@article{aaA_lib,
  author = {Wolf, Sarah E. and Liu, Tianyi and Govind, Shivajee and Zhao, Haoqiang and Huang, Georgia and Zhang, Aixi and Wu, Yu and Chin, Jocelyn and Cheng, Kevin and Salami-Ranjbaran, Elmira and Gao, Feng and Gao, Gui and Jin, Yi and Pu, Youge and Toledo, Thiago Gomes and Ablajan, Keyume and Walsh, Patrick J. and Fakhraai, Zahra},
  title = {Design of a homologous series of molecular glassformers},
  journal = {The Journal of Chemical Physics},
  volume = {155},
  number = {22},
  pages = {224503},
  year = {2021},
  month = {12},
}

@article{aaA_syn,
  author = {Liu, Tianyi and Cheng, Kevin and Salami-Ranjbaran, Elmira and Gao, Feng and Glor, Ethan C. and Li, Mu and Walsh, Patrick J. and Fakhraai, Zahra},
  title = {Synthesis and high-throughput characterization of structural analogues of molecular glassformers: 1{,}3{,}5-trisarylbenzenes},
  journal = {Soft Matter},
  volume = {11},
  pages = {7558-7566},
  year = {2015},
}

@article{bagchi_origin_2019,
  author = {Bagchi, Kushal and Jackson, Nicholas E. and Gujral, Ankit and Huang, Chengbin and Toney, Michael F. and Yu, Lian and de Pablo, Juan J. and Ediger, M. D.},
  title = {Origin of Anisotropic Molecular Packing in Vapor-Deposited Alq3 Glasses},
  journal = {The Journal of Physical Chemistry Letters},
  publisher = {American Chemical Society},
  volume = {10},
  number = {2},
  pages = {164--170},
  year = {2019},
  month = jan,
  doi = {10.1021/acs.jpclett.8b03582},
}

@article{bagchi_over_2020,
  author = {Bagchi, Kushal and Deng, Chuting and Bishop, Camille and Li, Yuhui and Jackson, Nicholas E. and Yu, Lian and Toney, M. F. and de Pablo, J. J. and Ediger, M. D.},
  title = {Over What Length Scale Does an Inorganic Substrate Perturb the Structure of a Glassy Organic Semiconductor?},
  journal = {{ACS} Applied Materials \& Interfaces},
  publisher = {American Chemical Society},
  volume = {12},
  number = {23},
  pages = {26717--26726},
  year = {2020},
  month = jun,
  doi = {10.1021/acsami.0c06428},
}

@article{berthier_origins_2017,
  author = {Berthier, Ludovic and Charbonneau, Patrick and Flenner, Elijah and Zamponi, Francesco},
  title = {Origin of Ultrastability in Vapor-Deposited Glasses},
  journal = {Phys. Rev. Lett.},
  publisher = {American Physical Society},
  volume = {119},
  pages = {188002},
  year = {2017},
  month = nov,
  doi = {10.1103/PhysRevLett.119.188002},
}

@article{bishop_sg_ratetempsuperposition,
  author = {Bishop, Camille and Li, Yuhui and Toney, Michael F and Yu, Lian and Ediger, M D},
  title = {Molecular Orientation for Vapor-Deposited Organic Glasses Follows Rate-Temperature Superposition: The Case of Posaconazole.},
  journal = {J Phys Chem B},
  volume = {124},
  number = {12},
  pages = {2505--2513},
  year = {2020},
  month = mar,
}

@article{bonomi_promoting_2019,
  author = {Bonomi, Massimiliano and Bussi, Giovanni and Camilloni, Carlo and Tribello, Gareth A. and Banáš, Pavel and Barducci, Alessandro and Bernetti, Mattia and Bolhuis, Peter G. and Bottaro, Sandro and Branduardi, Davide and Capelli, Riccardo and Carloni, Paolo and Ceriotti, Michele and Cesari, Andrea and Chen, Haochuan and Chen, Wei and Colizzi, Francesco and De, Sandip and De La Pierre, Marco and Donadio, Davide and Drobot, Viktor and Ensing, Bernd and Ferguson, Andrew L. and Filizola, Marta and Fraser, James S. and Fu, Haohao and Gasparotto, Piero and Gervasio, Francesco Luigi and Giberti, Federico and Gil-Ley, Alejandro and Giorgino, Toni and Heller, Gabriella T. and Hocky, Glen M. and Iannuzzi, Marcella and Invernizzi, Michele and Jelfs, Kim E. and Jussupow, Alexander and Kirilin, Evgeny and Laio, Alessandro and Limongelli, Vittorio and Lindorff-Larsen, Kresten and Löhr, Thomas and Marinelli, Fabrizio and Martin-Samos, Layla and Masetti, Matteo and Meyer, Ralf and Michaelides, Angelos and Molteni, Carla and Morishita, Tetsuya and Nava, Marco and Paissoni, Cristina and Papaleo, Elena and Parrinello, Michele and Pfaendtner, Jim and Piaggi, Pablo and Piccini, Giovanni Maria and Pietropaolo, Adriana and Pietrucci, Fabio and Pipolo, Silvio and Provasi, Davide and Quigley, David and Raiteri, Paolo and Raniolo, Stefano and Rydzewski, Jakub and Salvalaglio, Matteo and Sosso, Gabriele Cesare and Spiwok, Vojtěch and Šponer, Jiří and Swenson, David W.H. and Tiwary, Pratyush and Valsson, Omar and Vendruscolo, Michele and Voth, Gregory A. and White, Andrew},
  title = {Promoting transparency and reproducibility in enhanced molecular simulations},
  journal = {Nature Methods},
  volume = {16},
  number = {8},
  pages = {670--673},
  year = {2019},
  month = aug,
  doi = {10.1038/s41592-019-0506-8},
}

@article{brunel_in_2024,
  author = {Brunel, Fabrice and Galanopoulo, Paul and Espinosa Rodriguez, Edgar and Lansalot, Muriel and D'Agosto, Franck},
  title = {In situ SAXS investigation of vinyl acetate polymerization-induced self-assembly},
  journal = {Polymer Chemistry},
  volume = {15},
  number = {10},
  pages = {979-990},
  year = {2024},
  month = {03},
  doi = {10.1039/d3py01381b},
}

@article{chiu_coarse-grained_2010,
  author = {Chiu, See-Wing and Scott, H. Larry and Jakobsson, Eric},
  title = {A Coarse-Grained Model Based on Morse Potential for Water and N-Alkanes},
  journal = {Biophysical Journal},
  publisher = {Elsevier},
  volume = {98},
  number = {3},
  pages = {573a--574a},
  year = {2010},
  month = jan,
  doi = {10.1016/j.bpj.2009.12.3116},
}

@article{dalal_tunable_2015,
  author = {Shakeel S. Dalal  and Diane M. Walters  and Ivan Lyubimov  and Juan J. de Pablo  and M. D. Ediger},
  title = {Tunable molecular orientation and elevated thermal stability of vapor-deposited organic semiconductors},
  journal = {Proceedings of the National Academy of Sciences},
  volume = {112},
  number = {14},
  pages = {4227-4232},
  year = {2015},
  doi = {10.1073/pnas.1421042112},
}

@article{dodda_ligpargen_2017,
  author = {Dodda, Leela S. and Cabeza de Vaca, Israel and Tirado-Rives, Julian and Jorgensen, William L.},
  title = {LigParGen web server: an automatic OPLS-AA parameter generator for organic ligands},
  journal = {Nucleic Acids Research},
  volume = {45},
  number = {W1},
  pages = {W331-W336},
  year = {2017},
  month = {07},
  doi = {10.1093/nar/gkx312},
}

@article{dodda_localized_2017,
  author = {Dodda, Leela S. and Vilseck, Jonah Z. and Tirado-Rives, Julian and Jorgensen, William L.},
  title = {1.14*CM1A-LBCC: Localized Bond-Charge Corrected CM1A Charges for Condensed-Phase Simulations},
  journal = {The Journal of Physical Chemistry B},
  volume = {121},
  number = {15},
  pages = {3864-3870},
  year = {2017},
  month = {02},
  doi = {10.1021/acs.jpcb.7b00272},
}

@article{eastman_openmm_2024,
  author = {Eastman, Peter and Galvelis, Raimondas and Peláez, Raúl P. and Abreu, Charlles R. A. and Farr, Stephen E. and Gallicchio, Emilio and Gorenko, Anton and Henry, Michael M. and Hu, Frank and Huang, Jing and Krämer, Andreas and Michel, Julien and Mitchell, Joshua A. and Pande, Vijay S. and Rodrigues, João Pglm and Rodriguez-Guerra, Jaime and Simmonett, Andrew C. and Singh, Sukrit and Swails, Jason and Turner, Philip and Wang, Yuanqing and Zhang, Ivy and Chodera, John D. and De Fabritiis, Gianni and Markland, Thomas E.},
  title = {{OpenMM} 8: {Molecular} {Dynamics} {Simulation} with {Machine} {Learning} {Potentials}},
  journal = {The Journal of Physical Chemistry. B},
  volume = {128},
  number = {1},
  pages = {109--116},
  year = {2024},
  month = jan,
  doi = {10.1021/acs.jpcb.3c06662},
}

@article{gujral_anisotropic_2018,
  author = {Gujral, Ankit and Yu, Lian and Ediger, M. D.},
  title = {Anisotropic organic glasses},
  journal = {Current Opinion in Solid State and Materials Science},
  volume = {22},
  number = {2},
  pages = {49--57},
  year = {2018},
  doi = {10.1016/j.cossms.2017.11.001},
}

@article{gujral_structural_2015,
  author = {Gujral, Ankit and O'Hara, Katherine A. and Toney, Michael F. and Chabinyc, Michael L. and Ediger, M. D.},
  title = {Structural characterization of vapor-deposited glasses of an organic hole transport material with X-ray scattering},
  journal = {Chemistry of Materials},
  volume = {27},
  number = {9},
  pages = {3341--3348},
  year = {2015},
  doi = {10.1021/acs.chemmater.5b00583},
}

@incollection{hansen_theory_2013,
  author = {Jean-Pierre Hansen and Ian R. McDonald},
  title = {Chapter 11 - Molecular Liquids},
  booktitle = {Theory of Simple Liquids (Fourth Edition)},
  publisher = {Academic Press},
  edition = {Fourth Edition},
  pages = {455-510},
  year = {2013},
  address = {Oxford},
  doi = {https://doi.org/10.1016/B978-0-12-387032-2.00011-8},
}

@article{jorgensen_development_1996,
  author = {Jorgensen, William L. and Maxwell, David S. and Tirado-Rives, Julian},
  title = {Development and testing of the {OPLS} all-atom force field on conformational energetics and properties of organic liquids},
  journal = {Journal of the American Chemical Society},
  volume = {118},
  number = {45},
  pages = {11225--11236},
  year = {1996},
  doi = {10.1021/ja9621760},
}

@article{jorgensen_interpretable_2026,
  author = {Jorgensen, Christian and Lin, Arthur Y. and Vasavada, Rhushil and Cersonsky, Rose K.},
  title = {Interpretable visualizations of data spaces for classification problems},
  journal = {Machine Learning: Science and Technology},
  year = {2026},
  doi = {10.1088/2632-2153/ae466e},
}

@article{jorgensen_potential_2005,
  author = {William L. Jorgensen  and Julian Tirado-Rives},
  title = {Potential energy functions for atomic-level simulations of water and organic and biomolecular systems},
  journal = {Proceedings of the National Academy of Sciences},
  volume = {102},
  number = {19},
  pages = {6665-6670},
  year = {2005},
  doi = {10.1073/pnas.0408037102},
}

@article{kim_comprehensive_2022,
  author = {Changsub Kim and Chatchai Jarumaneeroj and Wonchalerm Rungswang and Kyeong Sik Jin and Moonhor Ree},
  title = {A comprehensive small angle X-ray scattering analysis on morphological structure of semicrystalline linear polymer in bulk state},
  journal = {Polymer},
  volume = {243},
  pages = {124610},
  year = {2022},
  doi = {https://doi.org/10.1016/j.polymer.2022.124610},
}

@article{li_small_2016,
  author = {Li, Tao and Senesi, Andrew J. and Lee, Byeongdu},
  title = {Small Angle X-ray Scattering for Nanoparticle Research},
  journal = {Chemical Reviews},
  volume = {116},
  number = {18},
  pages = {11128-11180},
  year = {2016},
  doi = {10.1021/acs.chemrev.5b00690},
  note = {PMID: 27054962},
}

@article{liew_coarse_2003,
  author = {Chee Chin Liew and Masuhiro Mikami},
  title = {A coarse-grained model for particle dynamics simulations of complex fluids},
  journal = {Chemical Physics Letters},
  volume = {368},
  number = {3},
  pages = {346-351},
  year = {2003},
  doi = {https://doi.org/10.1016/S0009-2614(02)01856-0},
}

@article{liu_birefringent_2017,
  author = {Liu, Tianyi and Exarhos, Annemarie L. and Alguire, Ethan C. and Gao, Feng and Salami-Ranjbaran, Elmira and Cheng, Kevin and Jia, Tiezheng and Subotnik, Joseph E. and Walsh, Patrick J. and Kikkawa, James M. and Fakhraai, Zahra},
  title = {Birefringent Stable Glass with Predominantly Isotropic Molecular Orientation},
  journal = {Phys. Rev. Lett.},
  publisher = {American Physical Society},
  volume = {119},
  pages = {095502},
  year = {2017},
  month = aug,
  doi = {10.1103/PhysRevLett.119.095502},
}

@article{lyubimov_orientational_2015,
  author = {Lyubimov, Ivan and Antony, Lucas and Walters, Diane M. and Rodney, David and Ediger, M. D. and de Pablo, Juan J.},
  title = {Orientational anisotropy in simulated vapor-deposited molecular glasses},
  journal = {The Journal of Chemical Physics},
  volume = {143},
  number = {9},
  pages = {094502},
  year = {2015},
  month = sep,
  doi = {10.1063/1.4928523},
}

@article{madanchi_is_2025,
  author = {Madanchi, Ata and Azek, Emna and Zongo, Karim and Béland, Laurent K. and Mousseau, Normand and Simine, Lena},
  title = {Is the Future of Materials Amorphous? Challenges and Opportunities in Simulations of Amorphous Materials},
  journal = {{ACS} Physical Chemistry Au},
  publisher = {American Chemical Society},
  volume = {5},
  number = {1},
  pages = {3--16},
  year = {2025},
  month = jan,
  doi = {10.1021/acsphyschemau.4c00063},
}

@article{pedersen_analysis_1997,
  author = {Jan Skov Pedersen},
  title = {Analysis of small-angle scattering data from colloids and polymer solutions: modeling and least-squares fitting},
  journal = {Advances in Colloid and Interface Science},
  volume = {70},
  pages = {171-210},
  year = {1997},
  doi = {https://doi.org/10.1016/S0001-8686(97)00312-6},
}

@article{rodriguez_surface_2019,
  author = {Rodr\'{\i}guez-Tinoco, Cristian and Gonzalez-Silveira, Marta and R\`afols-Rib\'e, Joan and Vila-Costa, Ana and Martinez-Garcia, Julio Cesar and Rodr\'{\i}guez-Viejo, Javier},
  title = {Surface-Bulk Interplay in Vapor-Deposited Glasses: Crossover Length and the Origin of Front Transformation},
  journal = {Phys. Rev. Lett.},
  publisher = {American Physical Society},
  volume = {123},
  pages = {155501},
  year = {2019},
  month = oct,
  doi = {10.1103/PhysRevLett.123.155501},
}

@article{sg_physics,
  author = {Berthier, Ludovic and Ediger, Mark D.},
  title = {Facets of glass physics},
  journal = {Physics Today},
  volume = {69},
  number = {1},
  pages = {40-46},
  year = {2016},
}

@article{shirts_statistically_2008,
  author = {Shirts, Michael R. and Chodera, John D.},
  title = {Statistically optimal analysis of samples from multiple equilibrium states},
  journal = {Journal of Chemical Physics},
  volume = {129},
  number = {12},
  pages = {124105--124105},
  year = {2008},
  doi = {10.1063/1.2978177},
}

@article{siachouli_effect_2024,
  author = {Siachouli, Panagiota and Karadima, Katerina S. and Mavrantzas, Vlasis G. and Pandis, Spyros N.},
  title = {The effect of functional groups on the glass transition temperature of atmospheric organic compounds: a molecular dynamics study},
  journal = {Soft Matter},
  volume = {20},
  number = {24},
  pages = {4783-4794},
  year = {2024},
  month = {06},
  doi = {10.1039/d4sm00405a},
}

@article{suter_rapid_2025,
  author = {Suter, James L. and M\:uller, Werner A. and Vassaux, Maxime and Anastasiou, Alexandros and Simmons, Martin and Tilbrook, David and Coveney, Peter V.},
  title = {Rapid, Accurate and Reproducible Prediction of the Glass Transition Temperature Using Ensemble-Based Molecular Dynamics Simulation},
  journal = {Journal of Chemical Theory and Computation},
  volume = {21},
  number = {3},
  pages = {1405-1421},
  year = {2025},
  month = {01},
  doi = {10.1021/acs.jctc.4c01364},
}

@article{swallen_organic_2007,
  author = {Swallen, Stephen F. and Kearns, Kenneth L. and Mapes, Marie K. and Kim, Yong Seol and McMahon, Robert J. and Ediger, M. D. and Wu, Tian and Yu, Lian and Satija, Sushil},
  title = {Organic {Glasses} with {Exceptional} {Thermodynamic} and {Kinetic} {Stability}},
  journal = {Science},
  publisher = {American Association for the Advancement of Science},
  volume = {315},
  number = {5810},
  pages = {353--356},
  year = {2007},
  month = jan,
  doi = {10.1126/science.1135795},
}

@article{thompson_lammps_2022,
  author = {Thompson, Aidan P. and Aktulga, H. Metin and Berger, Richard and Bolintineanu, Dan S. and Brown, W. Michael and Crozier, Paul S. and in 't Veld, Pieter J. and Kohlmeyer, Axel and Moore, Stan G. and Nguyen, Trung Dac and Shan, Ray and Stevens, Mark J. and Tranchida, Julien and Trott, Christian and Plimpton, Steven J.},
  title = {{LAMMPS} - a flexible simulation tool for particle-based materials modeling at the atomic, meso, and continuum scales},
  journal = {Computer Physics Communications},
  volume = {271},
  pages = {108171},
  year = {2022},
  month = feb,
  doi = {10.1016/j.cpc.2021.108171},
}

@article{thoms_in_2020,
  author = {Thoms, E. and Gabriel, J. P. and Guiseppi-Elie, A. and Ediger, M. D. and Richert, R.},
  title = {In situ observation of fast surface dynamics during the vapor-deposition of a stable organic glass},
  journal = {Soft Matter},
  volume = {16},
  number = {48},
  pages = {10860-10864},
  year = {2020},
  month = {12},
  doi = {10.1039/d0sm01916j},
}

@article{tribello_plumed_2014,
  author = {Tribello, Gareth A. and Bonomi, Massimiliano and Branduardi, Davide and Camilloni, Carlo and Bussi, Giovanni},
  title = {{PLUMED} 2: {New} feathers for an old bird},
  journal = {Computer Physics Communications},
  volume = {185},
  number = {2},
  pages = {604--613},
  year = {2014},
  month = feb,
  doi = {10.1016/j.cpc.2013.09.018},
}

@article{waasmaier_new_1995,
  author = {Waasmaier, D. and Kirfel, A.},
  title = {New analytical scattering-factor functions for free atoms and ions},
  journal = {Acta Crystallographica Section A Foundations of Crystallography},
  volume = {51},
  number = {3},
  pages = {416--431},
  year = {1995},
  month = may,
  doi = {10.1107/S0108767394013292},
}

@article{walters_influence_2017,
  author = {Walters, Diane M. and Antony, Lucas and de Pablo, Juan J. and Ediger, M. D.},
  title = {Influence of Molecular Shape on the Thermal Stability and Molecular Orientation of Vapor-Deposited Organic Semiconductors},
  journal = {The Journal of Physical Chemistry Letters},
  volume = {8},
  number = {14},
  pages = {3380--3386},
  year = {2017},
  month = jul,
  doi = {10.1021/acs.jpclett.7b01097},
}

@misc{white_kdepy_2023,
  author = {Tommy and                   Scott White and                   Alexandra and                   Riccardo Di Maio and                   gunvor and                   Luke Dyer},
  title = {tommyod/KDEpy: v1.1.8},
  publisher = {Zenodo},
  year = {2023},
  month = oct,
  doi = {10.5281/zenodo.8422073},
}

@article{zhang_parameterizing_2014,
  author = {Na Zhang and Peng Zhang and Wei Kang and Danny Bluestein and Yuefan Deng},
  title = {Parameterizing the Morse potential for coarse-grained modeling of blood plasma},
  journal = {Journal of Computational Physics},
  volume = {257},
  pages = {726-736},
  year = {2014},
  doi = {https://doi.org/10.1016/j.jcp.2013.09.040},
}

@article{zhang_study_2007,
  author = {Jian Zhang and Yu Liang and Jizhong Yan and Jianzhong Lou},
  title = {Study of the molecular weight dependence of glass transition temperature for amorphous poly(l-lactide) by molecular dynamics simulation},
  journal = {Polymer},
  volume = {48},
  number = {16},
  pages = {4900-4905},
  year = {2007},
  doi = {https://doi.org/10.1016/j.polymer.2007.06.030},
}

@article{lin_expanding_2024,
    author = {Lin, Arthur and Huguenin-Dumittan, Kevin K. and Cho, Yong-Cheol and Nigam, Jigyasa and Cersonsky, Rose K.},
    title = {Expanding density-correlation machine learning representations for anisotropic coarse-grained particles},
    journal = {The Journal of Chemical Physics},
    volume = {161},
    number = {7},
    pages = {074112},
    year = {2024},
    month = {08},
    issn = {0021-9606},
    doi = {10.1063/5.0210910},
    url = {https://doi.org/10.1063/5.0210910},
}

@article{hoang_free_2011,
  title = {Free surface effects on thermodynamics and glass formation in simple monatomic supercooled liquids},
  author = {Hoang, V. V. and Dong, T. Q.},
  journal = {Phys. Rev. B},
  volume = {84},
  issue = {17},
  pages = {174204},
  numpages = {11},
  year = {2011},
  month = {Nov},
  publisher = {American Physical Society},
  doi = {10.1103/PhysRevB.84.174204},
  url = {https://link.aps.org/doi/10.1103/PhysRevB.84.174204}
}

@article{singh_molecular_2011,
    author = {Singh, Sadanand and de Pablo, Juan J.},
    title = {A molecular view of vapor deposited glasses},
    journal = {The Journal of Chemical Physics},
    volume = {134},
    number = {19},
    pages = {194903},
    year = {2011},
    month = {05},
    issn = {0021-9606},
    doi = {10.1063/1.3586805},
    url = {https://doi.org/10.1063/1.3586805},
}

@article{jikang_applicatioin_2025,
title = {The application of organic materials used in OLED display},
journal = {Results in Optics},
volume = {21},
pages = {100856},
year = {2025},
issn = {2666-9501},
doi = {https://doi.org/10.1016/j.rio.2025.100856},
url = {https://www.sciencedirect.com/science/article/pii/S2666950125000847},
author = {Liu Jikang},
}

@article{yokoyama_molecular_2011,
    author = {Yokoyama, Daisuke},
    title = {Molecular orientation in small-molecule organic light-emitting diodes},
    journal = {Journal of Materials Chemistry},
    volume = {21},
    number = {48},
    pages = {19187-19202},
    year = {2011},
    month = {12},
    issn = {0959-9428},
    doi = {10.1039/c1jm13417e},
    url = {https://doi.org/10.1039/c1jm13417e},
    eprint = {https://pubs.rsc.org/jm/article-pdf/21/48/19187/2412710/c1jm13417e.pdf},
}

@article{ferron_characterization_2022,
    author = {Ferron, Thomas J. and Thelen, Jacob L. and Bagchi, Kushal and Deng, Chuting and Gann, Eliot and de Pablo, Juan J. and Ediger, M. D. and Sunday, Daniel F. and DeLongchamp, Dean M.},
    title = {Characterization of the Interfacial Orientation and Molecular Conformation in a Glass-Forming Organic Semiconductor},
    journal = {ACS Applied Materials \& Interfaces},
    volume = {14},
    number = {2},
    pages = {3455-3466},
    year = {2022},
    month = {01},
    issn = {1944-8244},
    doi = {10.1021/acsami.1c19948},
    url = {https://doi.org/10.1021/acsami.1c19948},
    eprint = {https://pubs.acs.org/aamick/article-pdf/14/2/3455/12806651/am1c19948.pdf},
}

@article{
bishop_vapor_2019,
author = {Camille Bishop  and Jacob L. Thelen  and Eliot Gann  and Michael F. Toney  and Lian Yu  and Dean M. DeLongchamp  and M. D. Ediger },
title = {Vapor deposition of a nonmesogen prepares highly structured organic glasses},
journal = {Proceedings of the National Academy of Sciences},
volume = {116},
number = {43},
pages = {21421-21426},
year = {2019},
doi = {10.1073/pnas.1908445116},
URL = {https://www.pnas.org/doi/abs/10.1073/pnas.1908445116},
eprint = {https://www.pnas.org/doi/pdf/10.1073/pnas.1908445116},
}

@article{alahe_kinetically_2026,
    author = {Alahe, A. T. M. Mahbub and Ferron, Thomas J. and Bishop, Camille E.},
    title = {Kinetically arrested phase separation leads to tunable domain structures in vapor-deposited glasses},
    journal = {The Journal of Chemical Physics},
    volume = {165},
    number = {12},
    pages = {124511},
    year = {2026},
    month = {09},
    issn = {0021-9606},
    doi = {10.1063/5.0350268},
    url = {https://doi.org/10.1063/5.0350268},
}

@article{luo_surface_2023,
   author = "Luo, Peng and Fakhraai, Zahra",
   title = "Surface-Mediated Formation of Stable Glasses", 
   journal= "Annual Review of Physical Chemistry",
   year = "2023",
   volume = "74",
   number = "Volume 74, 2023",
   pages = "361-389",
   doi = "https://doi.org/10.1146/annurev-physchem-042018-052708",
   url = "https://www.annualreviews.org/content/journals/10.1146/annurev-physchem-042018-052708",
   publisher = "Annual Reviews",
   issn = "1545-1593",
   type = "Journal Article",
  }

@article{
moore_effects_2019,
author = {Alex R. Moore  and Georgia Huang  and Sarah Wolf  and Patrick J. Walsh  and Zahra Fakhraai  and Robert A. Riggleman },
title = {Effects of microstructure formation on the stability of vapor-deposited glasses},
journal = {Proceedings of the National Academy of Sciences},
volume = {116},
number = {13},
pages = {5937-5942},
year = {2019},
doi = {10.1073/pnas.1821761116},
URL = {https://www.pnas.org/doi/abs/10.1073/pnas.1821761116},
eprint = {https://www.pnas.org/doi/pdf/10.1073/pnas.1821761116},
}

@article{tylinski_vapor_2016,
    author = {Tylinski, M. and Chua, Y. Z. and Beasley, M. S. and Schick, C. and Ediger, M. D.},
    title = {Vapor-deposited alcohol glasses reveal a wide range of kinetic stability},
    journal = {The Journal of Chemical Physics},
    volume = {145},
    number = {17},
    pages = {174506},
    year = {2016},
    month = {11},
    issn = {0021-9606},
    doi = {10.1063/1.4966582},
    url = {https://doi.org/10.1063/1.4966582},
}

@article{samanta_exploring_2019,
    author = {Samanta, Subarna and Huang, Georgia and Gao, Gui and Zhang, Yue and Zhang, Aixi and Wolf, Sarah and Woods, Connor N. and Jin, Yi and Walsh, Patrick J. and Fakhraai, Zahra},
    title = {Exploring the Importance of Surface Diffusion in Stability
of Vapor-Deposited Organic Glasses},
    journal = {The Journal of Physical Chemistry B},
    volume = {123},
    number = {18},
    pages = {4108-4117},
    year = {2019},
    month = {05},
    issn = {1520-6106},
    doi = {10.1021/acs.jpcb.9b01012},
    url = {https://doi.org/10.1021/acs.jpcb.9b01012},
    eprint = {https://pubs.acs.org/jpcbfk/article-pdf/123/18/4108/11081314/jp9b01012.pdf},
}

@article{
chai_direct_2014,
author = {Y. Chai  and T. Salez  and J. D. McGraw  and M. Benzaquen  and K. Dalnoki-Veress  and E. Raphaël  and J. A. Forrest },
title = {A Direct Quantitative Measure of Surface Mobility in a Glassy Polymer},
journal = {Science},
volume = {343},
number = {6174},
pages = {994-999},
year = {2014},
doi = {10.1126/science.1244845},
URL = {https://www.science.org/doi/abs/10.1126/science.1244845},
eprint = {https://www.science.org/doi/pdf/10.1126/science.1244845},
}

\end{document}